\documentstyle[aps,amsfonts,epsfig]{revtex}
{\makeatletter%
\gdef\labeleqs#1{{%
\edef\@currentlabel{%
\ifappendixon\appletter\fi
\ifsecnumbers\ifnum\c@secnum>0
\arabic{secnum}.\fi\fi\arabic{equation}}
\label{#1}%
}}%
}

\twocolumn
\begin{document}
\newcommand{\beq}{\begin{eqnarray}}
\newcommand{\eeq}{\end{eqnarray}}
\newcommand{\diff}{\mbox{d}}
\newcommand{\pardis}{\langle \mu \rangle}
\newcommand{\mathz}{\mathbb{Z}}
\newcommand{\proj}{\vec{\Phi}(x)}
\newcommand{\vproj}{\hat{\Phi}(x)}
\newcommand{\latproj}{{f}^{a,b}(\vec{n}+\hat{\imath},t)}
\newcommand{\blatproj}{{f}^{a,b}(\vec{n},t)}
%
\draft
\widetext
\title{Colour confinement and dual superconductivity of the vacuum - II}

\author{A. Di Giacomo$^{a,c,1}$, B. Lucini$^{b,c,2}$,
L. Montesi$^{b,c,2}$, G. Paffuti$^{a,c,1}$}

\address{$^a$ Dipartimento di Fisica dell'Universit\`a, Via Buonarroti
2 Ed. B, I-56127 Pisa, Italy}

\address{$^b$ Scuola Normale Superiore, Piazza dei Cavalieri 7, I-56126
Pisa, Italy}

\address{$^c$ INFN sezione di Pisa, Via Vecchia Livornese 1291, I-56010
S. Piero a Grado (Pi), Italy}

\address{$^1$ e-mail address: digiaco, paffuti@mailbox.difi.unipi.it}
\address{$^2$ e-mail address: lucini, montesi@cibs.sns.it}

\maketitle
\widetext
\begin{abstract}
The dual superconductivity of the vacuum in $SU(3)$ gauge theory is
investigated by constructing a disorder parameter which signals monopole
condensation in various abelian projections and by studying numerically on the
lattice its behaviour at finite temperature. We find that the vacuum is a dual
superconductor with respect to each $U(1)$ of the residual gauge group
after abelian projection independently of the abelian projection chosen.
Like in the $SU(2)$ case (discussed in a companion paper) a finite
size scaling analysis enables us to extract the indices of the phase
transition and our analysis is consistent with independent determinations.
\end{abstract}
\pacs{PACS numbers: 11.15.Ha, 12.38.Aw, 14.80.Hv, 64.60.Cn}
\setcounter{page}{1}
\narrowtext
\section{Introduction}
In a companion paper \cite{SU2}, which we will quote as I, we have
presented the basic ideas about confinement and dual superconductivity of
the ground state of gauge theories, and how they can be tested in $SU(2)$
gauge theory.
Monopoles exist in gauge theories, carrying a conserved magnetic charge.
We have defined a disorder parameter $\pardis$ detecting dual
superconductivity as spontaneous breaking of the $U(1)$ symmetry related to
magnetic charge conservation. $\pardis \ne 0$ signals that the ground state is
a superposition of states with different magnetic charge, a phenomenon which
is denoted as condensation and which implies dual
superconductivity under very general assumptions.

In $SU(2)$ a monopole species can be associated to any operator in the adjoint
representation, with a corresponding magnetic $U(1)$ symmetry. Condensation
can be numerically investigated for different monopole species,
in connection with confinement, by lattice simulation at finite temperature.
The main results of this investigation for $SU(2)$ were the following:
\begin{itemize}
\item monopoles defined by different abelian projections do condense in the
confined phase, or $\pardis \ne 0$;
\item at deconfinement $\pardis \to 0$;
\item a finite size scaling analysis allows to determine the critical index
$\nu$ of the correlation length, the critical $\beta $ and the index $\delta$
by which $\pardis \to 0$. The determination of $\nu$ agrees with the ones done
by other methods, and indicates a $2$nd order phase transition.
Also $\beta _C$ coincides within errors with the known values;
\item all the monopole species considered have a similar behaviour,
and show dual superconductivity.
\end{itemize}
~\vphantom{a}
\vskip8\baselineskip\noindent
Our conclusion was that confinement is an order-disorder transition. The
symmetry which characterizes the dual order is not fully understood, but for
sure the different $\pardis$'s are good disorder parameters.

In this paper we shall extend the analysis to $SU(3)$ gauge group.
The essentials are not changed with respect to $SU(2)$. Some formal
complications come from the coexistence of two monopole charges for each
abelian projection (Sect.~\ref{abelproj}).
We have performed a systematic numerical investigation, for different
abelian projections. Also for $SU(3)$ we find dual superconductivity
in all the abelian projections that we have considered, again indicating
that the guess of ref. \cite{'tHooft3} that all monopoles are physically
equivalent is correct. $\pardis$ looks, within errors, the same for the two
independent monopole charges of a given abelian projection. A finite
size scaling analysis shows that the transition is first order.
Numerical details and results are given in Sect.~\ref{results}.

Sect.~\ref{conclusions} contains some concluding remarks.
\section{The abelian projection. Conserved monopole charges}
\label{abelproj}
In analogy with the $SU(2)$ case we shall denote by
\beq
\label{eq:1}
\phi(x) = \sum_{i=1}^8 \phi^i(x) F^i
\eeq
the generic local operator in the
adjoint representation. $F^i = \lambda^i/2$, with $\lambda^i$
the Gell-Mann matrices. We shall assume $\phi$ hermitian, or $\phi^i$ real in
any configuration.

It will be convenient to use the notation
\beq
\phi(x) = \vec{\phi}(x) \cdot \vec{F}
\eeq
for eq.~(\ref{eq:1}) and for any two operators $\phi_1$, $\phi_2$
\beq
2 \mbox{tr} \phi_1 \phi_2 = \vec{\phi}_1 \cdot \vec{\phi}_2 \equiv
\sum_{i=1}^8
\phi^i _1 \phi^i _2 \ .
\eeq

Any $\phi (x)$ can be diagonalized by a unitary transformation $U(x)$
\beq
U(x) \phi(x) U^{\dagger}(x) = \phi_D (x) \ .
\eeq
In the usual representation of the $\lambda$ matrices
\beq
\label{eq:5}
\phi_D = \varphi_D^a F^a + \varphi_D^b F^b \ ,
\eeq
where $F^a$ and $F^b$ are independent linear combinations of $F^3$ and $F^8$.
We shall choose
\beq
\label{eq:6}
F^a = F^8 \ , \qquad F^b = \frac{\sqrt{3}}{2} F^3 + \frac{F^8}{2}
\eeq
for reasons which will be clear below.

We now define \beq \label{eq:7} f ^a(x) &=& \left(U (x)
\right)^{\dagger}F ^a U(x) \ ,\\ \nonumber f ^b(x) &=& \left(U (x)
\right)^{\dagger}F ^b U(x) \ . \eeq $U(x)$ is defined as the
matrix which diagonalizes $\phi(x)$. To eliminate ambiguities the
eigenvalues can be ordered in decreasing order. $U(x)$ is
determined up to an arbitrary matrix $U_D(x)$ on the left $U(x)
\simeq U_D(x) U(x)$, with $U_D= \exp\left( i \alpha F^8 + i \beta
F^3 \right)$, i.e. up to a residual $U(1)^2$. From
eq.~(\ref{eq:5}) and~(\ref{eq:7}), in the usual representation of
Gell-Mann matrices \beq \phi_D = \frac{1}{2 \sqrt{3}}\left(
\begin{array}{c c c}
\varphi_D^a + 2 \varphi_D^b & 0 & 0\\
0 & \varphi_D^a - \varphi_D^b & 0\\
0 & 0 & - 2 \varphi_D^a - \varphi_D^b
\end{array} \right) \ ,
\eeq
with $\varphi_D^a \ge 0$, $\varphi_D^b \ge 0$.

The gauge transform $U(x)$ is singular at the sites where either
$\varphi_D^b = 0$, and
\beq
\phi_D = \frac{1}{2 \sqrt{3}} \varphi_D^a \left(
\begin{array}{r r r}
1 & 0 & 0\\
0 & 1 & 0\\
0 & 0 & - 2
\end{array} \right) \ ,
\eeq
or $\varphi_D^a = 0$, and
\beq
\phi_D = \frac{1}{2 \sqrt{3}} \varphi_D^b \left(
\begin{array}{r r r}
2 & 0 & 0\\
0 & -1 & 0\\
0 & 0 & - 1
\end{array} \right) \ .
\eeq
In both cases $\phi_D$ has two equal eigenvalues.

The two field tensors
\beq
\label{eq:2.1}
F_{\mu \nu} ^{a,b} = \frac{1}{2} \mbox{tr} \left(
{f}^{a,b}  \cdot {G}_{\mu \nu} - \frac{i}{g}
{f}^{a,b} \cdot \right.\\
\nonumber \left. \frac{4}{3} \left[ D_{\mu} {f}^{a,b} \ , \ D_{\nu} {f}^{a,b}
\right] \right)
\eeq
are the analogous of the 't Hooft's tensor \cite{'tHooft1} in $SU(2)$.
Like in $SU(2)$ the bilinear terms in $A_{\mu} A_{\nu}$ cancel.
In the abelian projected gauge $f^a(x)=F^a$ and $f^b(x)=F^b$ are $x$
independent, apart from singularities, and therefore in the domain in which
$U(x)$ is regular
\beq
F^{a,b}_{\mu \nu} =  {\partial}_{\mu} A^{a,b} _{\nu} - {\partial}_{\nu}
A_{\mu}^{a,b} \ .
\eeq

The cancellation of the bilinear term $A_{\mu} A_{\nu}$ between the two terms
of eq.~(\ref{eq:2.1}) is not automatic in $SU(3)$ for arbitrary choice of
$f^a$, $f^b$, as it was in $SU(2)$, and only works if $f^a$, $f^b$ belong to
$U(1)$ in the breaking $SU(3) \to SU(2) \times U(1)$, which is the case for
the choice of (\ref{eq:6}). Also the choice $\frac{\sqrt{3}}{2} F^3 -
\frac{1}{2} F^8$ for $f^a$ or $f^b$ would be legitimate.

As in $SU(2)$ $F^{a,b \star}_{\mu \nu}$, the dual tensor
to $F^{a,b}_{\mu \nu}$, define two magnetic currents $\partial ^{\nu}
F^{a,b\star}_{\mu \nu} = j^{a,b}_{\mu}$, which are conserved. The theory
has two conserved magnetic charges, $M^a,\ M^b$. Monopoles exist
at the points where $U(x)$ is singular: where $\varphi_D^b = 0$ the monopole
field is directed as $F^3 = \mbox{diag}( \ 1/2\ ,-1/2 \ ,0 \ )$,
where $\varphi_D^a$ is zero it is directed as $F^{3a} = -(1/2) F^3 +
(\sqrt{3}/2) F^8 = \mbox{diag}(\ 0\ ,1/2 \ ,-1/2 \ )$.
As in $SU(2)$ we shall investigate the invariance of the ground state
with respect to these magnetic $U(1)$'s, in connection with confinement.

On the lattice, we shall define the abelian projected fields as follows.
In the abelian projected representation we write the generic
link $U_{\mu}(n)$ in the form
\beq
\label{eq:11}
U_{\mu}(n) = e^{i \vec{V}_{\bot} \cdot \vec{F}_{\bot}}
e^{i(V_a F^a + V_b F^b)} \ ,
\eeq
with $\vec{F}_{\bot}$ a superposition of generators belonging to non zero
eigenvalues of the Cartan algebra.

Eq. (\ref{eq:11}) is easy to prove: it is a trivial consequence of the
Baker-Hausdorff formula. Like for $SU(2)$, the abelian part of a product
is the sum of the abelian parts of the factors, to order $a^2$
($a$ is the lattice spacing). The abelian magnetic fluxes through plaquettes,
one for each $U(1)$, can be defined, and are
identically conserved. The disorder parameter is again
\beq
\langle \mu ^{a,b} \rangle = \frac{Z[S + \Delta^{a,b} S]}{Z[S]} \ ,
\eeq
\beq
\Delta^{a,b} S = \frac{\beta}{3} \sum _{\vec{n}}
\mbox{Tr} \left\{ \Pi _{i0}(\vec{n},t) - \Pi _{i0} ^{\prime{a,b}} (\vec{n},t)
\right\} \ ,
\eeq
$\Pi _{i0}^{\prime{a,b}}(\vec{n},t)$ is obtained from $\Pi _{i 0} (\vec{n},t) =
U_i(\vec{n},t)U_0(\vec{n}+ \hat \imath,t)
\left(U_i (\vec{n},t+1)\right)^{\dag}\left(U_0(\vec{n},t)\right)^{\dag}$
by the change
\beq
U_i(\vec{n},t) &\to& e^{- i \Lambda (\vec{n},\vec{y}) \blatproj}
U_i(\vec{n},t)\\
\nonumber
&&e^{i A_{\bot i}^M (\vec{n},\vec{y}) \latproj}
e^{i \Lambda (\vec{n}+\hat{\imath},\vec{y}){f}^{a,b}
(\vec{n}+ \hat{\imath},t) } \,
\eeq
where
\beq
\vec{A}^M(\vec{n},\vec{y}) = \vec{A}^M_{\bot} (\vec{n},\vec{y}) +
\vec{\nabla} \Lambda (\vec{n},\vec{y}) \
\eeq
is the vector potential produced by a monopole.
The proof that $\langle \mu ^{a,b} \rangle$
creates a monopoles of the corresponding type is exactly
the same as for $SU(2)$.

Also for $SU(3)$ instead of $\langle \mu ^{a,b} \rangle$ it is convenient
to determine
\beq
\label{eq:3.25}
\rho ^{a,b} =  \frac{\diff}{\diff \beta} \log \langle \mu ^{a,b} \rangle
= \langle S \rangle _S - \langle S + \Delta S \rangle _{S + \Delta S}
\eeq
as a function of $\beta$. We do that on an asymmetric lattice
$N_s ^3 \times N_t$ ($N_s \gg N_t$ ) which provides the static
thermal equilibrium  at $T = 1/a(\beta)$.

The deconfining transition is known and has been studied using the Polyakov
loop order parameter \cite{Fukujita}. We will investigate if going from
deconfined to confined phase monopoles do condense to produce dual
superconductivity. $\langle \mu ^{a,b} \rangle$ will be the disorder
parameters.
\section{Numerical results}
\label{results} We determine the temperature dependence of $\rho$
on a lattice $N_s^3 \times N_t$ ($N_s \gg N_t$), with $N_t = 4$
and $N_s$ ranging from 12 to 32. For the reason discussed in
\cite{SU2}, we use periodic boundary conditions in the spatial
directions and $C^{\star}$-boundary conditions \cite{Wiese} in the
time direction. As in $SU(2)$ we diagonalize an operator ${\cal
O}$ belonging to the group ${\cal O} = \exp(i \Phi^a\lambda^a/2)$
and we identify $\Phi^a$ by ordering the imaginary part of
eigenvalues in decreasing order.

As for the $SU(2)$ case, we study the following projections:
\begin{itemize}
\item ${\cal O}$ is connected to the Polyakov line
$L(\vec{n},t) = \Pi _{t^{\prime}=t}^{N_t -1}
U_0(\vec{n},t^{\prime}) \Pi _{t^{\prime}=0}^{t-1}
U_0(\vec{n},t^{\prime})$ in the following way: \beq
\label{eq:4.29} {\cal O}(\vec{n},t) = \Pi _{t^{\prime}=t}^{N_t -1}
U_0(\vec{n},t^{\prime}) L^{\star} (\vec{n},0)
\Pi_{t^{\prime}=0}^{t -1} U_0(\vec{n},t^{\prime}) \ ; \eeq
(Polyakov projection on a $C^{\star}$-periodic lattice);
\item ${\cal O}$ is an open plaquette, i.e. a parallel transport on an elementary
square of the lattice \beq \label{eq:4.30} {\cal O}(n)&=&
\Pi_{ij}(n)\\ \nonumber &=& U_i(n)U_j(n+\hat{\imath})
\left(U_i(n+\hat{\jmath})\right) ^{\dag} \left(U_j(n)
\right)^{\dag}\ ; \eeq
\item ${\cal O}$ is the ``butterfly'' $F$
\beq \label{eq:4.31}
{\cal O}(n) &=&  F (n) \\ \nonumber &=&
U_x(n)U_y(n+\hat{x}) \left(U_x(n+\hat{y})\right)^{\dag} \left(
U_y(n)\right)^{\dag}\\ \nonumber
&&U_z(n)U_t(n+\hat{z})\left(U_z(n+\hat{t})\right)^{\dag}
\left(U_t(n)\right) ^{\dag} \ . \eeq
\end{itemize}
The trace of $F$ is the density of topological charge.

The simulation was done on a 128-node APE Quadrics Machine. We use an
overrelaxed heat-bath algorithm to compute the Wilson term of
eq.~(\ref{eq:3.25}), and a mixed algorithm as described in our
previous paper \cite{SU2} for the monopole term $\langle S + \Delta S
\rangle _{S + \Delta S}$.
Far from the critical region at each $\beta$ we sampled
over 4000 termalized configurations, each of them taken after 4 sweeps.
The errors have been obtained by using a Jackknife method on binned data,
as discussed in \cite{SU2}.
In the critical region a higher statistics is required. Typically the Wilson
term is more noisy. Thermalization was checked by monitoring the action
density and the probability distribution of the trace of the Polyakov loop.
The number of measurements was at least $300 \tau _C$, where $\tau _C$
is the correlation time of the considered set of data.

For $\vec{A}_{\bot}$ we use the Wu-Yang's parameterization; we have also
checked numerically that Dirac's form gives similar results, as expected.

In terms of $\rho$
\beq
\label{eq:4.33}
\langle \mu ^{a,b} \rangle = \exp \left( \int _0^{\beta} \rho^{a,b}
(\beta^{\prime})\diff \beta ^{\prime} \right) \ .
\eeq
Eq. (\ref{eq:4.33}) implies that if the dual $U(1)$ symmetry defined by
some abelian projection and by some abelian generator of the gauge
group is related to colour confinement, in the
thermodynamic limit $N_s \to \infty$ the corresponding
$\rho$ stays finite in the strong coupling region ($\beta < \beta_C$)
and goes to $- \infty$ linearly with $N_s$ in the weak coupling region
($\beta > \beta_C$). In the critical region, the abrupt decline
of $\pardis$ is signaled by a sharp negative peak of $\rho$; the value
of $\rho$ in this region must behave as a function of $N_s$ as prescribed
by the finite size scaling theory of pseudocritical systems.

\begin{figure}[t]
\begin{center}
\epsfig{figure=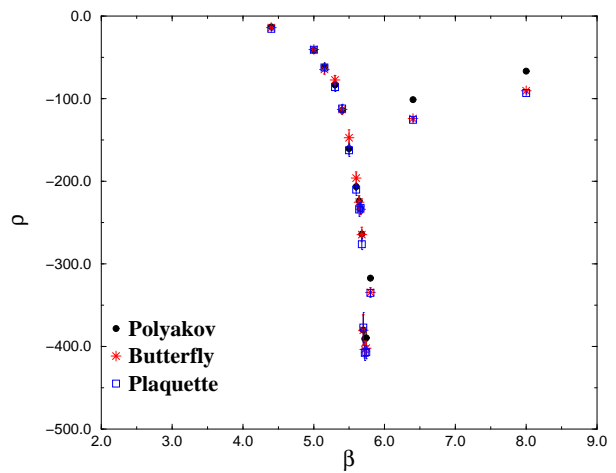, angle=270, width=8cm}
\end{center}
\caption{$\rho$ vs. $\beta$ for different abelian projections.
Lattice $12^3 \times 4$, abelian generator $F^3$.}
\label{condtot12.fig}
\null\vskip 0.3cm
\end{figure}

Fig.~\ref{condtot12.fig} shows the typical behaviour of $\rho$ for
different abelian projections, for a lattice $12^3 \times 4$.
As abelian generator we used $F^3$.
The negative peak occurs at the expected transition point, $\beta_C$
\cite{Karsch}. Below $\beta_C$ the different projections are equal
within errors, suggesting that different monopoles behave in the same
way.

\begin{figure}[htb]
\begin{center}
\epsfig{figure=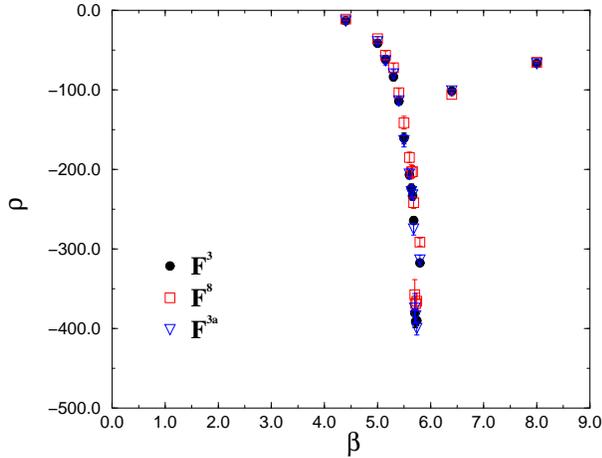, angle=0, width=8cm}
\end{center}
\caption{$\rho$ vs. $\beta$ for different abelian generators.
Lattice $12^3 \times 4$, Polyakov projection.}
\label{abeliangenerators.fig}
\null\vskip 0.3cm
\end{figure}

We have investigated also whether at fixed abelian projection the profile
of $\rho$ depends on the $U(1)$ magnetic subgroup.
Fig~\ref{abeliangenerators.fig} shows
the profile of $\rho$ corresponding to $F^3, F^8$ and $F^{3a}$
in the Polyakov projection on a $12^3 \times 4$ lattice. No appreciable
differences can be seen between different choices. This is an indication
(confirmed also by simulations on larger lattices) that monopoles
defined with respect to different abelian generators behave in the
same way in the $SU(3)$ vacuum. This is also true for the other
abelian projections we have investigated (see fig.~\ref{condtot12.fig}).

\begin{figure}[htb]
\begin{center}
\epsfig{figure=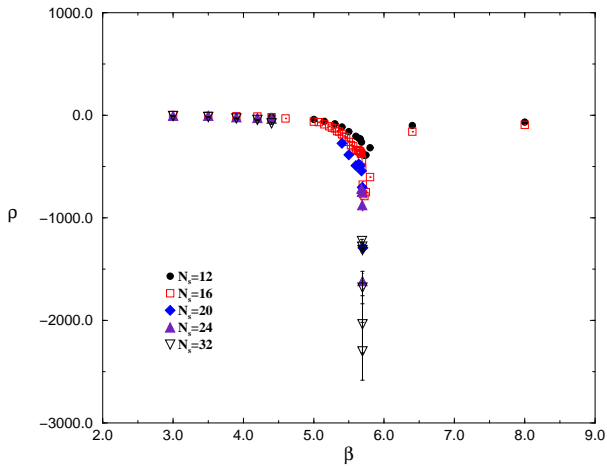, angle=270, width=8cm}
\end{center}
\caption{$\rho$ as a function of $\beta$ for different spatial sizes
at fixed $N_t=4$. Polyakov projection, abelian generator $F^3$.}
\label{rhopla16.fig}
\null\vskip 0.3cm
\end{figure}
Since different abelian projections and different abelian generators
give indistinguishable results, for the sake of
simplicity we shall only display the Polyakov projection and the abelian
generator $F^3$ in the following figures.

Fig.~\ref{rhopla16.fig} shows the dependence of $\rho$ on $N_s$.
The qualitative behaviour does not change when we enlarge
the lattice size.

We now analyze the dependence on $N_s$ in more detail.

\begin{figure}[htb]
\begin{center}
\epsfig{figure=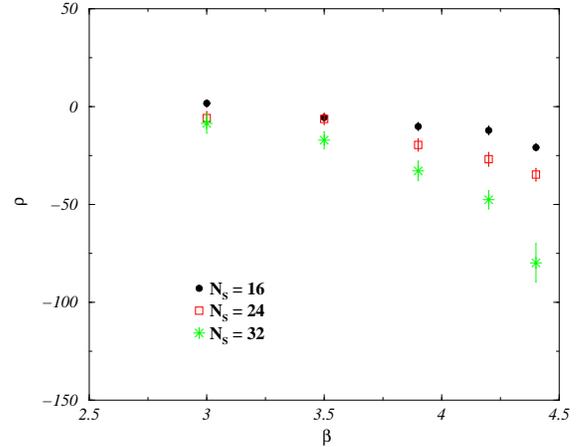, angle=0,width=7.5cm}
\end{center}
\caption{$\rho$ vs. $\beta$ in the strong coupling region for lattice
sizes $N_s^3 \times 4$. Polyakov projection, abelian generator $F^3$.}
\label{rhostrong.fig}
\null\vskip 0.8cm
\end{figure}

In the strong coupling region at low $\beta$'s $\rho$
seems to converge to a finite value (cfr. fig.~\ref{rhostrong.fig}).
Eq. (\ref{eq:4.33}) then implies that $\pardis \ne 0$ in the
infinite volume limit in the confined phase for these $\beta$'s.
Hence monopoles do condense in this phase.

\begin{figure}[htb]
\begin{center}
\epsfig{figure=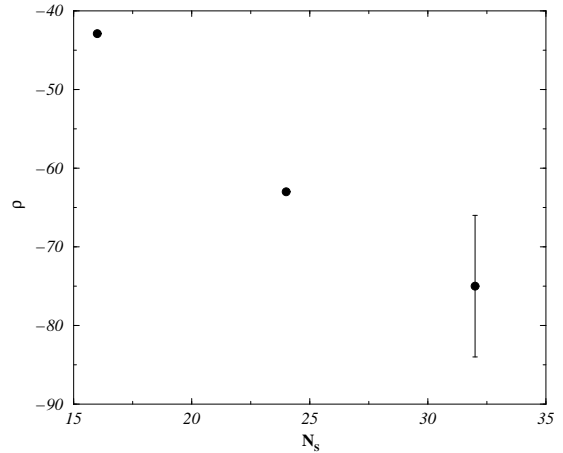, angle=0, width=7.35cm}
\end{center}
\caption{$\rho$~vs.~$N_s$ ($N_t = 4$) at $\beta = \infty$ in the
Polyakov projection with abelian generator $F^3$.~Data are obtained
by numerical minimization of $\langle S + \Delta S \rangle _{S + \Delta S}$.}
\label{rhoweak_su2.fig}
\null\vskip 0.3cm
\end{figure}

In the weak coupling region, we can evaluate $\rho$ perturbatively.
The path integral is then dominated by the classical
solutions of the equations of motion for the gauge variables
and we have
\beq
\rho
&&\mathop{\to}_{\beta \to \infty}
\left[ \mathop{\min}_{U} \{ S \} -
 \mathop{\min}_{U} \{ S+ \Delta S \} \right]\\
\nonumber
&& \ = - \mathop{\min}_{U} \{ S +
\Delta S \} \ ,
\eeq
since $\mathop{\min}_{U} \{ S \} = 0$.

In other systems, where the same shifting procedure has been applied
and studied, this asymptotic value has been analytically
calculated in perturbation
theory with the result \cite{CE97,Digiau1}
\beq\label{weakdec}
\rho = - c N_s + d \ ,
\eeq
where $c$ and $d$ are  constants, i.e $\rho$ goes linearly with
the spatial dimension.

In $SU(3)$ we are unable to perform the same calculation and
we have evaluated the minimum $\mathop{\min}_{U} \{ S + \Delta S \} $
numerically. Detail about the followed procedure have been discussed
in \cite{SU2}. Here we note that due to the single
precision of the APE Quadrics Machine,
the estimation of the minimum of $S + \Delta S$ for the biggest
lattice is more noisy than in the $SU(2)$ case.

The result is shown in fig.~\ref{rhoweak_su2.fig} for the Polyakov projection.
It is consistent with the linear dependence of eq. (\ref{weakdec})
with $c \simeq 2$ and $d \simeq -12$.
Thus in the weak coupling region in the thermodynamic limit $\rho$ goes
to $- \infty$ linearly with the spatial lattice size and
\beq
\langle \mu \rangle \mathop{\approx}_{N_s \to \infty}
A e^{ (- c N_s + d )\beta } \to 0,\quad \beta > \beta_C \ .
\eeq
The magnetic $U(1)$ symmetry is indeed restored in the deconfined phase.

The behaviour of $\rho$ in the critical region can be investigated
by using finite size scaling techniques. We know that the transition is weak
first order with a behaviour which is difficult to distinguish from that
of a second order transition.

By dimensional argument
\beq
\pardis = N_s ^{- \delta / \nu}
\Phi \left(\frac{\xi}{N_s},\frac{a}{\xi},\frac{N_t}{N_s} \right) \ ,
\eeq
where $a$ and $\xi$ are respectively the lattice spacing
and the correlation length of the system.

Near the critical point, for $\beta < \beta_C$
\beq
\xi \propto \left( \beta _C - \beta \right)^{- \nu} ,
\eeq
where $\nu$ is some effective critical exponent. In the limit
$N_s \gg N_t$  and for $a / \xi \ll 1$, i.e. sufficiently
close to the critical point we obtain
\beq
\pardis = N_s ^{- \delta / \nu}
\Phi \left( N_s^{1/\nu}\left( \beta _C - \beta \right),0,0 \right)
\eeq
or equivalently
\beq
\label{scaling}
\frac{\rho}{N_s^{1/\nu}} = f\left( N_s^{1/\nu} \left( \beta _C - \beta
\right)\right) \ .
\eeq
The ratio $\rho / N_s^{1/\nu}$ is a universal function of the
scaling variable
\beq
x =   N_s^{1/\nu} \left( \beta _C - \beta \right) \ .
\eeq
For a pseudocritical behaviour, we expect $\nu = 1/3$. Using also
$\beta_C(N_t=4) = 5.6925$ \cite{Karsch}, we can plot
$\rho / N_s^{1/\nu}$ as a function of $x$.

If we perform such a plot, we find that the scaling relation (\ref{scaling})
does not hold. Such a scaling violation is due to finite size
effect. A relationship more appropriate than (\ref{scaling})
is
\beq
\label{scaling2}
\frac{\rho}{N_s^{1/\nu}} = f\left( N_s^{1/\nu} \left( \beta _C - \beta
\right)\right) + \Phi(N_s) \ ,
\eeq
where $\Phi(N_s)$ parameterizes finite size effects. If we
assume that these effects are not critical \cite{Berg} then $\Phi$ is given
by
\beq
\Phi(N_s) = \frac{a}{N_s^3} \ ,
\eeq
where $a$ is a constant. This parameterization is correct ${\cal O}(1/N_s^6)$.

\begin{figure}[tb]
\begin{center}
\epsfig{figure=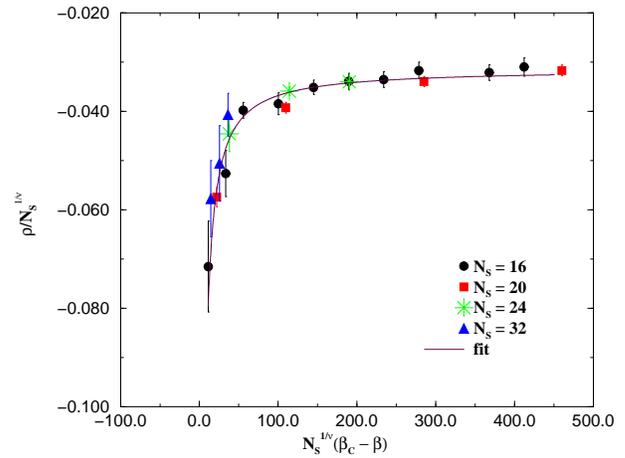, angle=0, width=8cm}
\end{center}
\caption{Quality of scaling in the Polyakov projection at $N_t=4$.
Abelian generator $F^3$.}
\label{scalapla.fig}
\null\vskip 0.3cm
\end{figure}

Fig. \ref{scalapla.fig} shows the quality of the scaling for $a=190$.
Our estimate gives $\nu = 0.33 \pm 0.07$ and $a = 190
\pm 20$.

In the thermodynamic limit in some region of $\beta < \beta_C$
we expect
\beq
\pardis \propto  \left( \beta _C - \beta \right) ^{\delta} \ ,
\eeq
which implies
\beq
\label{rhoas}
\frac{\rho}{N_s^{1/\nu}} = - \frac{\delta}{x}  \ .
\eeq

Using eq. (\ref{rhoas}) it should be possible in principle to determine
$\nu$, $\delta$ and $\beta _C$. Our statistic is not enough
accurate to perform such a fit. However, we can determine $\delta$
using as an input $\beta_C$, $\nu$, which are known, by
parameterizing $\rho$ in a wide range by the form
\beq
\frac{\rho}{N_s^{1/\nu}} = - \frac{\delta}{x}  - c  + \frac{a}{N_s^3} \ ,
\eeq
where $c$ is a constant, as suggested by fig.~\ref{scalapla.fig}.

Our best fit\footnote{Fits have been performed by using the
Minuit routines.} $\delta = 0.54 \pm 0.04$ for the Polyakov projection
and compatible results for the other projections. The $\chi ^2$ is
order 1.

This concludes our argument about the thermodynamic limit
($N_s \to \infty$). The deconfining
phase transition can be seen from a dual point of view as the
transition of the vacuum from the dual superconductivity phase
to the dual ordinary phase. That feature seems to be independent
of the abelian projection and of the abelian generator chosen.
\section{Concluding remarks}
\label{conclusions}
Like for $SU(2)$, also for $SU(3)$ gauge theory we have found evidence that
transition to deconfinement is an order-disorder transition, the disorder
parameter being a condensate of magnetic charges. A finite size scaling
analysis of the system gives critical indices compatible with a first order
transition, in agreement with determinations done by other methods
\cite{Fukujita}.

Of course we have investigated a limited number of abelian projections: like
in $SU(2)$, however, the indication is that physics is independent
of that choice.

An interesting issue would be to investigate if the mechanism is the same
in the $N_c = \infty$ limit. As a consequence also in the presence of dynamical
quarks the behaviour should be similar, as well as the symmetry pattern
and the disorder parameter should be the same. Investigation in this
direction is on the way.
\section*{Acknowledgements}
This work is partially supported by EC contract FMRX-CT97-0122 and
by MURST.

\end{document}